\documentclass{ws-procs9x6}
\def\today{31~December~2005} 

\def\LHS{lhs~}
\def \D {\hbox{d}}
\def \Log {\mathop{\rm Log}\nolimits}

\def \cotanh {\mathop{\rm cotanh}\nolimits}
\def \cotan {\mathop{\rm cotan}\nolimits}

\def \Im  {\mathop{\rm Im}\nolimits}
\def \arg {\mathop{\rm arg}\nolimits}
\def \mod#1{\vert #1 \vert}
\def \jmax{J}
\def\JPSJ{J.~Phys.~Soc.~Japan}
\def\PTP{Prog.~Theor.~Phys.~}

\def \bfE {{\bf E}}

\def \bfu {{\bf u}}

\begin{document}

\title{Integration of partially integrable equations}

\author{R.~CONTE}

\address{Service de physique de l'\'etat condens\'e (CNRS URA no.~2464)
\\~~CEA--Saclay, F--91191 Gif-sur-Yvette Cedex, France
\\ E-mail: Robert.Conte@cea.fr}

\maketitle

\abstracts{
Most evolution equations 
are partially integrable
and, in order to explicitly integrate all possible cases,
there exist several methods of complex analysis, but none is optimal.
The theory of Nevanlinna and Wiman-Valiron
on the growth of the meromorphic solutions
gives predictions and bounds,
but it is not constructive and restricted to meromorphic solutions.
The Painlev\'e approach via the \textit{a priori} singularities
of the solutions
gives no bounds but it is often (not always) constructive.
It seems that an adequate combination of the two methods
could yield much more output in terms of
explicit (i.e.~closed form) analytic solutions.
We review this question,
mainly taking as an example the chaotic equation of
Kuramoto and Sivashinsky
$\nu u''' + b u'' + \mu u' + u^2/2 +A=0, \nu \not=0$,
with $(\nu,b,\mu,A)$ constants.
}

\section{Introduction}
\label{sectionIntroduction}

Phenomena in continuous media
are often
governed by a partial differential equation (PDE),
e.g.~in one space variable $x$ and one time variable $t$
\begin{eqnarray}
& &
\bfE\left(
\left\lbrace
\frac{\partial_{m+n}}{\partial x^m \partial t^n} \bfu \right\rbrace
\right)=0,
\label{eqPDE}
\end{eqnarray}
in which $\bfu$ and $\bfE$ are multidimensional,
the integers $m,n$ take a finite set of values.
Our interest is the nonintegrable or even chaotic case,
for which the powerful tools of Lax pairs,
inverse spectral transform, etc \cite{AKNS} are inapplicable.
The derivation of analytic results must then use other methods.
Let us quote a few examples.

\begin{enumerate}
\item
\index{complex Ginzburg--Landau!(CGL3) equation}
The one-dimensional cubic complex Ginzburg-Landau equation (CGL3)
\begin{eqnarray}
{\hskip -10.0 truemm}
& &
i A_t + p A_{xx} + q \mod{A}^2 A - i \gamma A =0,\
p q \Im(p/q)\not=0,\
\label{eqCGL3}
\end{eqnarray}
(and its complex conjugate, i.e.~a total differential order four),
in which $p,q$ are complex constants and $\gamma$ a real constant,
a generic equation which describes many physical phenomena,
such as the propagation of a signal in an optical fiber
\cite{AgrawalBook},
spatiotemporal intermittency in spatially extended dissipative systems
\cite{MannevilleBook,vHSvS,vS2003}.
For two coupled CGL3 equations, see analytic results in Ref.~\cite{CM2000b}.

\item
The Kuramoto and Sivashinsky (KS) equation,
\index{Kuramoto--Sivashinsky (KS)!equation}
\begin{eqnarray}
{\hskip -10.0 truemm}
& &
\varphi_t + \nu \varphi_{xxxx} + b \varphi_{xxx} + \mu \varphi_{xx}
 + \varphi \varphi_x = 0,\
\nu \not=0,\
\label{eqKS}
\end{eqnarray}
in which $\nu,b,\mu$ are real constants.
This PDE is obeyed by the variable $\varphi=\arg A$ of the above field $A$
of CGL3 under some limit \cite{PM1979,Lega2001},
hence its name of phase turbulence equation.

\item
The quintic complex Ginzburg-Landau equation (CGL5),
\begin{eqnarray}
{\hskip -10.0 truemm}
& &
i A_t +p A_{xx} +q \mod{A}^2 A +r \mod{A}^4 A -i \gamma A =0,\
p r \Im(p/r)\not=0,\
\label{eqCGL5}
\end{eqnarray}
in which $p,q,r$ are complex constants and $\gamma$ a real constant.

\item
The Swift-Hohenberg equation \cite{SH1977,LMN}
\index{Swift-Hohenberg equation}
\begin{eqnarray}
{\hskip -10.0 truemm}
& &
i A_t +b A_{xxxx} +p A_{xx} +q \mod{A}^2 A +r \mod{A}^4 A -i \gamma A =0,\
b r \not=0,\
\label{eqSH}
\end{eqnarray}
in which $b,p,q,r$ are complex constants and $\gamma$ a real constant.

\end{enumerate}


The autonomous nature of (\ref{eqPDE})
(absence of any explicit dependence in $x$ and $t$)
allows the existence of
\textit{travelling waves} $u=U(\xi)$,
solutions of the ordinary differential equation (ODE)
\begin{eqnarray}
& &
u(x,t)=U(\xi),\ \xi=x-ct,\
E(U^{(N)},U^{(N-1)},\dots,U',U)=0.
\label{eqODEReduced}
\end{eqnarray}

For the CGL3, KS, CGL5 and Swift-Hohenberg equations
(with one exception, KS with $b^2=16 \mu \nu$),
all the solitary wave solutions
$\mod{A}^2=f(\xi), 
\varphi=\Phi(\xi), \xi=x-ct$,
which are known hitherto are polynomials in $\tanh k \xi$
(or $\cotanh,\tan,\cotan$, which are the same in the complex plane),
and such solutions are easy to find by taking advantage of the
singularity structure of the PDE
(see, e.g., the summer school lecture notes \cite{CetraroConte}).

Hence the natural questions: 
(i) Can other solitary waves $u=f(x-ct)$ exist (in closed form)?
(ii) If yes, please find them all, not just a few ones.

The present paper introduces to the methods
in principle able to answer both questions.
They will mainly be exemplified with the KS equation
(\ref{eqKS}).

The paper is organized as follows.
In section \ref{sectionFormulation},
we give a mathematical formulation of the problem.
In section \ref{sectionCount},
we prove the inexistence of an analytic expression
representing the \textit{general solution},
and we compute the gap
between
the differential order $N$ of the ODE (\ref{eqODEReduced})
and the maximal number of integration constants
in a singlevalued solution.
In section \ref{sectionMissingSolutions},
we give hints (not proofs) that some analytic result still has to be found.
In section \ref{sectionConsequencesSingle},
we review the consequences of the assumption of singlevaluedness
for a solution of the ODE (\ref{eqODEReduced}),
and present an algorithm to implement them.
In section \ref{sectionConsequencesMeromorphy},
we present the consequences of the assumption of meromorphy
for a solution of (\ref{eqODEReduced}).
The last section \ref{sectionSummary}
states the open problems.

\section{Mathematical formulation of the problem}
\label{sectionFormulation}

The successive steps of the announced program are
\begin{enumerate}
\item
To perform the traveling wave reduction from the PDE to an ODE.
The KS PDE (\ref{eqKS}) depends on three \textit{fixed} constants
$(\nu,b,\mu)$ (fixed means: which occur in the definition of the equation),
the reduction
\begin{eqnarray}
& &
\varphi(x,t)=c+u(\xi),\ \xi=x-ct,\
\end{eqnarray}
introduces in the ODE one more fixed constant $A$
(the second constant $c$ cancels out because of the Galilean invariance)
\begin{eqnarray}
& &
 \nu u''' + b u'' + \mu u' + \frac{u^2}{2} + A = 0,\ \nu \not=0,
\label{eqKSODE}
\end{eqnarray}
and the general solution of (\ref{eqKSODE}),
\textit{if it exists},
depends on the four fixed constants $(\nu,b,\mu,A)$
and three \textit{movable} constants
(movable means: which depends on the initial data),
which are the origin $\xi_0$ of $\xi$ and two other constants $c_1,c_2$.

\item
To count the number of constants which survive in the general solution of
(\ref{eqKSODE})
when one requires singlevaluedness.

\item
To find this largest singlevalued particular solution
in closed form.
Indeed, its representation as a series can be misleading,
as shown by classical authors like Poincar\'e and Painlev\'e.

\end{enumerate}

\section{Local separation of singlevaluedness and multivaluedness}
\label{sectionCount}

Because the ODE (\ref{eqODEReduced}) is nonintegrable,
the number of integration constants present in any
closed form solution
is strictly smaller than the differential order of the ODE.
This difference,
an indicator of the amount of integrability of the ODE,
can be precisely computed from a local analysis.

Two local representations of the general solution of
(\ref{eqODEReduced}) exist.
The first one, also the most well known,
is useless for our purpose.
This is the famous Taylor series near a regular point,
whose existence, unicity, convergence, etc is stated by
the existence theorem of Cauchy.
The reason why it is useless is its inability to make a distinction
between chaotic ODEs such as (\ref{eqKSODE})
and integrable  ODEs such as $u''' - 12 u u' -1=0$.

The second one, less known than the Taylor series of Cauchy,
is a Laurent series
(or more generally psi-series and/or Puiseux series)
near a movable singularity $x_0$.
This one does provide the expected information.
The technique to compute it is just the Painlev\'e test
(see Ref.~\cite{Cargese1996Conte} for the basic vocabulary of this technique).
Let us present it on the KS example (\ref{eqKSODE}).
\index{Kuramoto--Sivashinsky (KS)!equation}

Looking for a local algebraic behaviour
near a movable singularity $x_0$
\begin{eqnarray}
& &
u \sim_{x \to x_0} u_0 \chi^p,\ u_0 \not=0,\ \chi=x-x_0,
\end{eqnarray}
one first balances the highest derivative and the nonlinearity,
\begin{eqnarray}
& &
p-3=2p,\ p(p-1)(p-2) \nu u_0 + \frac{u_0^2}{2}=0,\
\label{eqKSLeadingOrder}
\end{eqnarray}
a system easily solved as
\begin{eqnarray}
& &
p=-3,\ u_0=120 \nu.
\label{eqKSLeadingOrderSol}
\end{eqnarray}
The resulting convergent Laurent series,
\begin{eqnarray}
& &
u^{(0)} = \frac{120 \nu}{\chi^{3}} - \frac{15 b}{\chi^{2}}
        + \frac{15 (16 \mu \nu - b^2)}{4 \times 19 \nu \chi}
        + \frac{13 (4  \mu \nu - b^2) b}{32 \times 19 \nu^2}
        + O(\chi^1),
\label{eqKSODELaurent}
\end{eqnarray}
lacks two of the three arbitrary constants.
They appear in perturbation \cite{CFP1993},
\begin{eqnarray}
& &
u= u^{(0)} + \varepsilon u^{(1)} + \varepsilon^2 u^{(2)} + \dots,
\end{eqnarray}
in which 
$\varepsilon$ is not in the ODE (\ref{eqKSODE}).
The linearized equation around $u^{(0)}$
\begin{eqnarray}
& &
\left(\nu \frac{\D^3}{\D x^3} + b \frac{\D^2}{\D x^2} + \mu \frac{\D}{\D x}
 + u^{(0)}\right) u^{(1)}=0,
\end{eqnarray}
has then the Fuchsian type near $x_0$,
with an indicial equation
($q=-6$ denotes the singularity degree of the \LHS\ of (\ref{eqKSODE}))
\begin{eqnarray}
& &
\lim_{\chi \to 0} \chi^{-j-q} (\nu \partial_x^3 + u_0 \chi^p) \chi^{j+p}
\\
& &
= \nu (j-3)(j-4)(j-5) + 120 \nu = \nu (j+1) (j^2 -13 j + 60).
\end{eqnarray}
The resulting local representation of the general solution,
\begin{eqnarray}
u(x_0,\varepsilon c_{-1},\varepsilon c_+,\varepsilon c_-)
& = &
 120\nu \chi^{-3} \{
 \hbox{Regular}(\chi)
\nonumber
\\
& &
+ \varepsilon [
              c_{-1} \chi^{-1} \hbox{Regular}(\chi)
\nonumber
\\
& &
        + \ \ c_{+} \chi^{(13+i\sqrt{71})/2} \hbox{Regular}(\chi)
\nonumber
\\
& &
        + \ \ c_{-} \chi^{(13-i\sqrt{71})/2} \hbox{Regular}(\chi) ]
            + {\mathcal O}(\varepsilon^2)\},
\label{eqKSGeneralSol}
\end{eqnarray}
in which ``Regular'' denotes converging series,
depends on 4 arbitrary constants
$(x_0,\varepsilon c_{-1},\varepsilon c_+,\varepsilon c_-)$
but, as shown by Poincar\'e,
the contribution of $\varepsilon c_{-1}$
is the derivative of
(\ref{eqKSODELaurent}) with respect to $x_0$,
so $c_{-1}$ can be set to zero.
The dense movable branching due to the irrational indices
reflects \cite{TF} the chaos,
and to remove it one has to require
$\varepsilon c_+=\varepsilon c_-=0$, i.e.~$\varepsilon=0$,
making the analytic part of (\ref{eqKSGeneralSol}) to depend on
the single arbitrary constant $x_0$.

The ODE (\ref{eqKSODE}) admits other Laurent series
in the variable $(u-\sqrt{-2 A})^{-1}$,
but they provide no additional information.

The question is then to turn this local information into a global one,
i.e.~to find the closed form singlevalued expression
depending on the maximal number (here one)
of movable constants.

We will call
\textit{unreachable}
 \index{unreachable}
any constant of integration which cannot participate to any
closed form solution.
The KS ODE (\ref{eqKSODE}) has two unreachable integration constants,
the third one $x_0$ being irrelevant since it reflects the
invariance of (\ref{eqKSODE}) under a translation of $x$.

We will also call
\textit{general analytic solution}
 \index{general analytic solution}
the closed form solution which depends on the maximal possible number
of reachable integration constants,
and our goal is precisely to exhibit a closed form expression
for this general analytic solution,
whose local representation is a Laurent series like
(\ref{eqKSODELaurent}).

The above notions (irrelevant, unreachable) belong to an
equation, not to a solution.
Let us  introduce another integer number,
attached to a solution,
allowing one to measure its distance to the general analytic solution.

The
\textit{distance}
 \index{distance}
of a closed form solution to the general analytic solution
is defined as the number of constraints
between the fixed constants and the reachable relevant constants.

For the ODE (\ref{eqKSODE}),
the fixed constants are $\nu,b,\mu,A$,
the movable constant $x_0$ is irrelevant,
the movable constants $c_1=\varepsilon c_+,c_2=\varepsilon c_-$ are unreachable,
so the distance $d$
is the number of constraints
among the fixed constants.

The closed form singlevalued solutions known to date are
\begin{enumerate}
\item
one elliptic solution (distance $d=1$)
\cite{FournierSpiegelThual,Kud1990}
\begin{eqnarray}
& &
{\hskip -18.0 truemm}
b^2 = 16 \mu \nu :\
u=-60\nu \wp' - 15b\wp - \frac{b \mu}{4 \nu},\
g_2=\frac{\mu^2}{12 \nu^2},\
g_3=\frac{13\mu^3+\nu A}{1080 \nu^3},
\label{eqKSConstraint16}
\label{eqKSElliptic}
\end{eqnarray}
in which $\wp$ is the elliptic function of Weierstrass,
\begin{eqnarray}
& &
{\wp'}^2=4 \wp^3 - g_2 \wp - g_3,
\label{eqWeierstrass}
\end{eqnarray}

\item
six trigonometric solutions ($d=2$)
\cite{KuramotoTsuzuki,KudryashovKSFourb},
rational in $e^{k \xi}$,
\begin{eqnarray}
u & = &
 120 \nu \tau^3
 - 15 b \tau^2
 +\left(\frac{60}{19} \mu - 30 \nu k^2 - \frac{15 b^2}{4 \times 19 \nu}\right)
 \tau
\nonumber
\\
& &
 + \frac{5}{2} b k^2 - \frac{13 b^3}{32 \times 19 \nu^2}
  + \frac{7 \mu b}{4 \times 19 \nu},\
\tau  =  \frac{k}{2} \tanh \frac{k}{2} (\xi-\xi_0),
\label{eqKSTrigo}
\end{eqnarray}
the allowed values being listed in Table \ref{TableKS},

\item
one rational solution ($d=3$),
\begin{eqnarray}
& &
b=0,\ \mu=0,\ A=0\ :\ u = 120\nu (\xi-\xi_0)^{-3},
\end{eqnarray}
which is a limit of all the previous solutions.
\end{enumerate}

\begin{table}[h] 
\caption[garbage]{
The six trigonometric solutions of KS, Eq.~(\ref{eqKSODE}).
They all have the form (\ref{eqKSTrigo}).
The last line is a degeneracy of the elliptic solution
(\ref{eqKSElliptic}).
}
\vspace{0.2truecm}
\begin{center}
\begin{tabular}{| c | c | c |}
\hline 
$b^2/(\mu\nu)$ & $\nu A/\mu^3$ & $\nu k^2/\mu$
\\ \hline \hline 
$0$ & $-4950/19^3,\ 450/19^3$ & $11/19,\ -1/19$
\\ \hline 
$144/47$ & $-1800/47^3$ & $1/47$
\\ \hline 
$256/73$ & $-4050/73^3$ & $1/73$
\\ \hline 
$16$ & $-18,\ -8$ & $1,\ -1$
\\ \hline 
\end{tabular}
\end{center}
\label{TableKS}
\end{table}

All those solutions admit the representation
\begin{eqnarray}
& &
{\hskip -8.0 truemm}
u=
\mathcal{D} \Log \psi + \hbox{constant},\
\mathcal{D}=60 \nu \frac{\D^3 }{\D \xi^3} + 15 b \frac{\D^2 }{\D \xi^2}
           + \frac{15(16 \mu \nu - b^2)}{76 \nu} \frac{\D }{\D \xi},
\label{eqKSD}
\end{eqnarray}
in which $\psi$ is an \textit{entire} function.
This linear operator $\mathcal{D}$,
which captures the singularity structure,
is called the
\textit{singular part operator}.
 \index{singular part operator}

The Laurent series (\ref{eqKSODELaurent}) yields another
information \cite{Hone2005}.
If its sum is elliptic,
the sum of the residues of the poles inside a period parallelogram must vanish.
Since the only poles of (\ref{eqKSODE}) are one triple pole,
a necessary condition \cite{Hone2005} for the sum to be elliptic is
to cancel the residue of (\ref{eqKSODELaurent}),
i.e. $b^2 = 16 \mu \nu$.
For this equation,
the condition is also sufficient, see (\ref{eqKSElliptic}).

\section{Experimental and numerical evidence of missing solutions}
\label{sectionMissingSolutions}

Experiments or computer simulations
display regular patterns in the $(x,t)$ plane
(see \cite{vS2003}),
some patterns being described by an analytic expression.
For the other patterns,
the guess is that there should exist matching analytic expressions.
For the equation (\ref{eqKS}),
one has observed a homoclinic wave \cite[Fig.~7]{Toh}
$\varphi=f(\xi),\xi=x-ct$,
while all known solutions are heteroclinic.
\index{Kuramoto--Sivashinsky (KS)!equation}

The Laurent series (\ref{eqKSODELaurent}) only provides a local knowledge
of the general analytic solution.
Rather than obtaining a global knowledge of the solution,
which is the ultimate goal,
it is easier to look at its singularities,
by computing the \textit{Pad\'e approximants} \cite{PadeIntro}
of the Laurent series (\ref{eqKSODELaurent}).
Pad\'e approximants are a powerful tool
to study the singularities of the sum
of a given Taylor series,
and more generally to perform the summation of divergent series.

Given the first $N$ terms of a Taylor series near $x=0$,
\begin{eqnarray}
& &
S_N = \sum\limits_{j=0}^N c_j x^j,
\end{eqnarray}
the Pad\'e approximant $[L,M]$ of the series is the unique rational
function
\begin{eqnarray}
& &
[L,M] = \frac{\sum_{l=0}^L a_l x^l}{\sum_{m=0}^M b_m x^m},\ b_0=1,
\end{eqnarray}
obeying the condition
\begin{eqnarray}
& &
S_N - [L,M] = {\mathcal O}(x^{N+1}),\qquad L+M=N.
\end{eqnarray}
The extension to Laurent series presents no difficulty.
In particular, for $L$ and $M$ large enough,
Pad\'e approximants are exact on rational functions .

The advantage of $\lbrack L,M \rbrack$ over $S_N$
(which has no poles)
is to display the \textit{global structure} of singularities of the series.

From a  thorough investigation \cite{YCM2003} of the singularities of the
sum of the Laurent series (\ref{eqKSODELaurent})
one concludes (this is not a proof):
for generic values of $(\nu,b,\mu,A)$,
no multivaluedness is detected,
no cuts are detected,
and the singularities look arranged in a nearly doubly periodic pattern,
the elementary cell being made of one triple pole and three simple zeroes.

\section{Consequences of singlevaluedness (Painlev\'e)}
\label{sectionConsequencesSingle}

\subsection{Classical results on first order autonomous equations}
\label{sectionClassicalResultsFirstOrderAutonomous}

The failure to detect any multivaluedness
in the unknown general analytic solution
by no means implies the singlevaluedness of this
general analytic solution,
because the Painlev\'e test only generates necessary conditions,
and the Pad\'e approximants are a numerical investigation.
It is however worthwhile to examine in detail the consequences of
an assumed singlevaluedness.

Given the $N$-th order autonomous algebraic ODE (\ref{eqODEReduced}),
any solution is
\begin{eqnarray}
& &
u = f(\xi-\xi_0),
\label{eqFormalSolution}
\end{eqnarray}
in which $\xi_0$ is movable.
Provided the elimination of $\xi_0$
between the equation (\ref{eqFormalSolution}) and its derivative
is possible,
one obtains the first order nonlinear ODE
\begin{eqnarray}
& &
F(u,u') =0,
\label{eqOrder1Autonomous}
\end{eqnarray}
in which $F$ is as unknown as $f$.

However, $f(\xi-\xi_0)$ is now
the \textit{general solution} of (\ref{eqOrder1Autonomous}),
and there exist classical results on first order autonomous
ODEs which are in addition \textit{algebraic}.
Let us therefore assume from now on
that the dependence of $f$ on $\xi_0$ is algebraic
(this is a sufficient condition for $F$ to be algebraic).

Let us summarize.
Given the $N$-th order ODE (\ref{eqODEReduced})
and
its particular solution $f$ Eq.~(\ref{eqFormalSolution}),
and assuming the dependence of $f$ on $\xi_0$ to be algebraic,
one is able to derive a first order ODE (\ref{eqOrder1Autonomous})
which is algebraic.

Conversely, given an algebraic first order ODE $F=0$
Eq.~(\ref{eqOrder1Autonomous}),
is it possible to go back to $f$?
This question has been answered positively by
Briot and Bouquet, Fuchs, Poincar\'e
and put in final form by Painlev\'e \cite[pages 58--59]{PaiLecons}.

\begin{theorem}
Given the algebraic first order ODE $F=0$
Eq.~(\ref{eqOrder1Autonomous}),
if its general solution is singlevalued, then
\begin{enumerate}

\item
Its general solution is an elliptic function, possibly degenerate,
and its expression is known in closed form.

\item
The genus of the algebraic curve (\ref{eqOrder1Autonomous}) is one or zero.

\item
There exist a positive integer $m$ and $(m+1)^2$
complex constants $a_{j,k}$, with $a_{0,m}\not=0$, such that
the polynomial $F$ has the form
\begin{eqnarray}
& &
F(u,u') \equiv
 \sum_{k=0}^{m} \sum_{j=0}^{2m-2k} a_{j,k} u^j {u'}^k=0,\ a_{0,m}\not=0.
\label{eqsubeqODEOrderOnePP}
\end{eqnarray}

\end{enumerate}
\label{Theorem1}
\end{theorem}

Then, assuming $f$
singlevalued with an algebraic dependence on $\xi_0$,
\begin{enumerate}

\item
It is equivalent to search for the solution $f$ or for $F$.

\item
The solution $f$ can only be elliptic
(i.e.~rational in $\wp$ and $\wp'$),
or a rational function of $e^{a x}$ with $a$ constant,
or a rational function of $x$.

\end{enumerate}

The explicit form
(\ref{eqsubeqODEOrderOnePP}) of $F$
makes it much easier to look for $F$ than $f$.

\subsection{Method to obtain the first order autonomous subequation}
\label{sectionClassNecessary}

The input data and assumptions are:
\begin{enumerate}
\item
a $N$-th order algebraic ODE (\ref{eqODEReduced}), $N \ge 2$,

\item
a Laurent series representing its general analytic solution,

\item
a first order algebraic ODE 
sharing its general solution with (\ref{eqODEReduced}).

\end{enumerate}

Then, by the classical results of section
\ref{sectionClassicalResultsFirstOrderAutonomous},
$F$ has the form (\ref{eqsubeqODEOrderOnePP}),
and there exists an algorithm \cite{MC2003}
yielding the solution $f$ in the canonical form
\begin{eqnarray}
& &
u=R(\wp',\wp)=R_1(\wp) + \wp' R_2(\wp),
\label{equWeierstrassForm}
\end{eqnarray}
in which $R_1,R_2$ are two rational functions,
with the possible degeneracies
\begin{eqnarray}
& &
R(\wp',\wp)
\longrightarrow
R(e^{k \xi})
\longrightarrow
R(\xi),
\end{eqnarray}
in which $R$ denotes rational functions.
This algorithm is \cite[Section 5]{MC2003}:

\begin{enumerate}

\item
Compute finitely many terms of the Laurent series,
\begin{eqnarray}
& &
u=\chi^p
 \left(\sum_{j=0}^{\jmax} u_j \chi^j+{\mathcal O}(\chi^{\jmax+1})\right),\
\chi=\xi-\xi_0.
\label{eqLaurent}
\end{eqnarray}

\item
Choose a positive integer $m$ and define the
first order ODE
\begin{eqnarray}
& &
F(u,u') \equiv
 \sum_{k=0}^{m} \sum_{j=0}^{[(m-k)(p-1)/p]} a_{j,k} u^j {u'}^k=0,\
a_{0,m}\not=0,
\label{eqsubeqODEOrderOnePPwithp}
\end{eqnarray}
in which $[z]$ denotes the integer part function.
The upper bound on $j$ implements the condition
$m(p-1) \le j p + k (p-1)$,
identically satisfied if $p=-1$,
that no term can be more singular than ${u'}^m$.

\item
Require the Laurent series to satisfy the Briot and Bouquet ODE,
i.e.~require the identical vanishing of the Laurent series for the \LHS\
$F(u.u')$ up to the order $\jmax$
\begin{eqnarray}
& &
F \equiv \chi^{m(p-1)} \left(\sum_{j=0}^{\jmax} F_j \chi^j
 + {\mathcal O}(\chi^{\jmax+1})
\right),\
\forall j\ : \ F_j=0.
\label{eqLinearSystemFj}
\end{eqnarray}
If it has no solution for $a_{j,k}$, increase $m$ and return to first step.

\item
For every solution,
integrate the first order autonomous ODE (\ref{eqsubeqODEOrderOnePPwithp}).
\end{enumerate}
The main step is to solve the set of equations (\ref{eqLinearSystemFj}),
i.e.~a linear, overdetermined system in the unknowns $a_{j,k}$.
This is quite an easy task.

An upper bound on $m$
will be established in section \ref{sectionConsequencesMeromorphy}.

\subsection{Results of the method on the KS equation}
\label{sectionKSsubeqFirstOrder}

\index{Kuramoto--Sivashinsky (KS)!equation}

The Laurent series of (\ref{eqKSODE}) is (\ref{eqKSODELaurent}).
In the second step, the smallest integer $m$ allowing a 
triple
pole ($p=-3$) in (\ref{eqsubeqODEOrderOnePPwithp}) is $m=3$.
With the normalization $a_{0,3}=1$, the subequation contains ten coefficients,
which are first determined by the Cramer system of ten equations
$F_j=0,j=0:6,8,9,12$.
The remaining overdetermined nonlinear system for $(\nu,b,\mu,A)$
contains as greatest common divisor (gcd) $b^2-16 \mu \nu$,
which defines a first solution
\begin{eqnarray}
& &
\frac{b^2}{\mu \nu}=16,\
u_s=u+\frac{3 b^3}{32 \nu^2},\
\nonumber
\\
& &
\left(u' + \frac{b}{2 \nu} u_s\right)^2
\left(u' - \frac{b}{4 \nu} u_s\right)
+\frac{9}{40 \nu}
\left(u_s^2 + \frac{15 b^6}{1024 \nu^4} + \frac{10 A}{3}\right)^2=0.
\label{eqKSsubeqgenus1}
\end{eqnarray}
After division by this gcd,
the remaining system for $(\nu,b,\mu,A)$
admits four solutions
(stopping the series at $\jmax=16$ is enough),
namely the first three lines of Table \ref{TableKS},
each solution defining a subequation,
\begin{eqnarray}
& &
{\hskip -10.0 truemm}
b=0,\
\nonumber
\\
& &
{\hskip -10.0 truemm}
\left(u' + \frac{180 \mu^2}{19^2 \nu}\right)^2
\left(u' - \frac{360 \mu^2}{19^2 \nu}\right)
+\frac{9}{40 \nu}
\left(u^2 + \frac{30 \mu}{19} u' - \frac{30^2 \mu^3}{19^2 \nu}\right)^2=0,\
\label{eqKSsubeqgenus0first}
\\
& &
{\hskip -10.0 truemm}
b=0,\
{u'}^3
+\frac{9}{40 \nu}
\left(u^2 + \frac{30 \mu}{19} u' + \frac{30^2 \mu^3}{19^3 \nu}\right)^2=0,\
\\
& &
{\hskip -10.0 truemm}
\frac{b^2}{\mu \nu}=\frac{144}{47},\
u_s=u-\frac{5 b^3}{144 \nu^2},\
\left(u' + \frac{b}{4 \nu} u_s\right)^3+\frac{9}{40 \nu} u_s^4=0,\
\\
& &
{\hskip -10.0 truemm}
\frac{b^2}{\mu \nu}=\frac{256}{73},\
u_s=u-\frac{45 b^3}{2048 \nu^2},\
\nonumber
\\
& &
{\hskip -10.0 truemm}
\left(u' + \frac{b}{8 \nu} u_s\right)^2
\left(u' + \frac{b}{2 \nu} u_s\right)
+\frac{9}{40 \nu}
\left(u_s^2+\frac{5 b^3}{1024 \nu^2}u_s + \frac{5 b^2}{128 \nu}u'\right)^2=0,\
\nonumber
\\
& &
{\hskip -10.0 truemm}
\label{eqKSsubeqgenus0last}
\end{eqnarray}

To integrate the subequations
(\ref{eqKSsubeqgenus1}),
(\ref{eqKSsubeqgenus0first})--(\ref{eqKSsubeqgenus0last}),
one first computes their genus
\footnote{For instance with the Maple command \textit{genus} of the
package \textit{algcurves} \cite{MapleAlgcurves},
which implements an algorithm of Poincar\'e.},
which is one for (\ref{eqKSsubeqgenus1}),
    and zero for (\ref{eqKSsubeqgenus0first})--(\ref{eqKSsubeqgenus0last}).
Therefore (\ref{eqKSsubeqgenus1}) has an elliptic general solution,
listed above as (\ref{eqKSElliptic}).
The general solution of the four others
(\ref{eqKSsubeqgenus0first})--(\ref{eqKSsubeqgenus0last})
is the third degree polynomial (\ref{eqKSTrigo})
in $\tanh k (\xi-\xi_0)/2$.

These four solutions,
obtained for the minimal choice of the subequation degree $m$,
constitute all the analytic results currently known on (\ref{eqKSODE}).

\section{Consequences of meromorphy (Nevanlinna)}
\label{sectionConsequencesMeromorphy}

If the solution $f$ is meromorphic,
much can be said from the study of its growth at infinity
(Nevanlinna theory).
For the KS ODE,
the meromorphy requires $c_+=c_-=0$ in
(\ref{eqKSGeneralSol}),
restricting the solution to the series
(\ref{eqKSODELaurent}).

By direct application of the Nevanlinna theory,
one can prove the
\begin{theorem}
\cite{EremenkoKS}
If a solution of (\ref{eqKSODE}) is meromorphic,
then it is elliptic or degenerate of elliptic. Furthermore,
\begin{enumerate}
\item
Elliptic solutions only exist if $b^2=16 \mu \nu$,
and their order is three. 

\item
Exponential solutions 
have the necessary form $P(\tanh k (\xi-\xi_0))$,
with $k$ constant and $P$ a polynomial of degree three.

\item
The only rational solution is $u=120 \nu (\xi-\xi_0)^{-3}$,
it exists for $b=\mu=A=0$.                                       

\end{enumerate}
\end{theorem}

Consequently,
the value $m=3$ is an upper bound to the algorithm of section
\ref{sectionClassNecessary},
which has therefore found all the meromorphic solutions
of (\ref{eqKSODE}).

\section{Summary and open problems}
\label{sectionSummary}

Let us represent the solutions of (\ref{eqKSODE}) by the
following inclusions,
\begin{eqnarray}
& &
\hbox{elliptic}
\subset
\hbox{meromorphic}
\subset
\hbox{singlevalued}
\subset
\hbox{multivalued}.
\end{eqnarray}

One has seen the various implications
\begin{enumerate}
\item
(Singlevalued, algebraic dependence on $x_0$)
$\Longrightarrow$
elliptic (thm \ref{Theorem1}), 

\item
Meromorphic
$\Longrightarrow$
elliptic (\cite{EremenkoKS} using Nevanlinna theory),

\item
Elliptic
$\Longrightarrow$
($b^2=16 \mu \nu$) (residue theorem \cite{Hone2005}),

\item
(Elliptic or degenerate)
$\Longrightarrow$
(order three) (\cite{EremenkoKS} using Nevanlinna theory)
$\Longrightarrow$
(all such solutions in closed form \cite{MC2003}).

\end{enumerate}

The problem is open
to find the general analytic solution in closed form
for arbitrary $(\nu,b,\mu,A)$,
which would be the sum of the Laurent series (\ref{eqKSODELaurent}).
Pad\'e approximants and Painlev\'e analysis find no multivaluedness nowhere.

Two and only two possibilities remain about this
general analytic solution
for generic values of $(\nu,b,\mu,A)$,
\begin{enumerate}
\item
either it is multivalued,
and strong efforts have then to be made to uncover this
multivaluedness with both the Painlev\'e test and the Pad\'e approximants.
This event is unlikely;

\item
or it is singlevalued.
In this case it cannot be elliptic,
and the dependence on $x_0$ is necessarily transcendental.

\end{enumerate}

Solving this open problem
would solve \textit{ipso facto} many similar problems for
nonintegrable equations such as CGL3, CGL5 or Swift-Hohenberg.

\section*{Acknowledgments}

The author warmly thanks the organizers for invitation.

\vfill \eject
\end{document}